\documentclass[a4paper,11pt]{article}
\usepackage{pos}

\usepackage{graphicx}
\usepackage{subcaption}

\begin{document}
\title{Gluon nonlocal operator mixing in
lattice QCD}

\author*{Demetrianos Gavriel}
\author{Haralambos Panagopoulos}
\author{Gregoris Spanoudes}

\affiliation{University of Cyprus,\\
  1 Panepistimiou Avenue, 2109 Aglantzia, Cyprus}

\emailAdd{gavriel.demetrianos@ucy.ac.cy}
\emailAdd{panagopoulos.haris@ucy.ac.cy}
\emailAdd{spanoudes.gregoris@ucy.ac.cy}

\abstract{
In this study, we explore the renormalization of a comprehensive set of gauge-invariant gluon nonlocal operators on the lattice. We calculate the renormalization factors for these operators in the modified Minimal Subtraction $(\rm \overline{MS})$ scheme up to one-loop, using both dimensional and lattice regularizations in the Wilson gluon action. To facilitate a non-perturbative renormalization approach, we examine an appropriate version of the modified regularization-invariant (${\rm RI}'$) scheme and determine the conversion factors from this scheme to $\rm \overline{MS}$. As an integral part of this procedure, by employing symmetry arguments on the lattice, we identify the mixing pattern of these operators under renormalization.
\begin{center}
\includegraphics[scale=0.45]{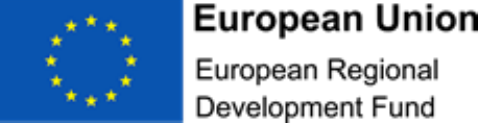}
\includegraphics[scale=0.45]{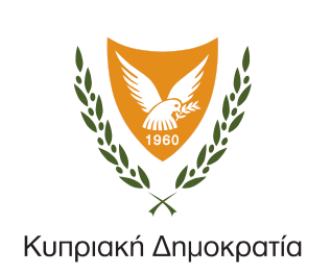}
\includegraphics[scale=0.45]{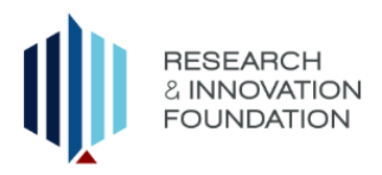}
\includegraphics[scale=0.45]{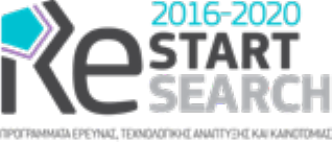}
\end{center}
}

\FullConference{The 41st International Symposium on Lattice Field Theory (LATTICE2024)\\
 28 July - 3 August 2024\\
Liverpool, UK\\}

\maketitle

\section{Introduction}

Quark Parton Distribution Functions (PDFs) have been widely studied, but research on gluon PDFs has been more limited. Nonetheless, gluons play a crucial role in understanding key physical phenomena, such as the proton's spin \cite{Alexandrou:2017oeh, Alexandrou:2020sml, Yang:2018bft}. Accurate calculations of gluon-dependent quantities are essential for processes like Higgs boson production, heavy quarkonium, and jet production \cite{Butterworth:2015oua, Sato:2019yez, Hou:2016nqm, Harland-Lang:2014zoa}. Moreover, gluon PDFs dominate over quark PDFs in the small-$x$ region, as observed in phenomenological studies \cite{Alekhin2014}. Therefore, a robust framework for first-principles calculations of gluon PDFs using lattice QCD is necessary to complement experimental research.

Lattice QCD methods developed to extract the x-dependence of quark distributions are also applied to gluon PDFs. These methods include the study of quasi-PDFs \cite{Wang:2017qyg, Wang2019, Zhang:2018diq, Fan:2018dxu} and pseudo-ITDs \cite{Balitsky:2019krf, Fan:2020cpa, Fan:2021bcr}, where a thorough understanding of their nonperturbative renormalization is essential. Lattice studies on quark quasi-PDFs \cite{Constantinou2017} reveal both linear and logarithmic divergences in the Wilson-line operator matrix elements, as well as mixing among certain subsets of operators under renormalization. The renormalization of gluon PDFs is expected to face similar challenges. Recently, it has been demonstrated that different components of nonlocal gluon operators have nontrivial renormalization patterns by employing the auxiliary field approach \cite{Zhang2018}. Additional relevant studies can be found in Refs.~\cite{Dorn:1981wa, Wang:2017qyg, Wang:2017eel, Wang:2019tgg, Braun:2020ymy}.

In these proceedings, we summarize the key points of the original paper \cite{Gavriel:2024qii}, omitting technical details; please refer to the paper for those. We present the one-loop renormalization factors for gluon nonlocal Wilson-line operators in the \(\overline{\rm MS}\) scheme, analyze their symmetry properties and mixing patterns, and provide conversion factors to the RI\(^\prime\) scheme.

\section{Theoretical setup}

\subsection{Definition of nonlocal gluon operators}

The nonlocal gluon operators in the fundamental representation are expressed as:
\begin{equation}
    O_{\mu \nu \rho \sigma} (x+ z \hat{\tau},x) \equiv 2 \ {\rm Tr} \bigg( F_{\mu \nu}(x+z \hat{\tau}) W(x+z \hat{\tau},x) F_{\rho \sigma}(x) W(x, x+z \hat{\tau})  \bigg)
    \label{eq:nonlocal_operator}
\end{equation}
Here, \( F_{\mu \nu} \) is the gluon field strength tensor, and \( W(x, x + z \hat{\tau}) \) represents the straight Wilson line of length \( z \). It is expressed as the path-ordered (\(\mathcal{P}\)) exponential of the gauge field \( A_\mu \), given by:
\begin{equation}
   W(x,x+z \hat{\tau}) \equiv \mathcal{P} \, \exp{\left[ i g \int_0 ^z A_\mu (x+\zeta \hat{\tau}) \, d\zeta\right]}
\end{equation}
Henceforth, without loss of generality, we consider the case where \( x = 0 \), with the origin positioned at one endpoint of the operator. Additionally, we set \( \tau = 3 \), so that the Wilson line lies along the \( z \)-direction. Under this fixed choice, the antisymmetry of $F_{\mu\nu}$ leads to 36 nonlocal gluon operators, determined by all possible index combinations of $O_{\mu \nu \rho \sigma}$.

Nonlocal gluon operators may generally mix under renormalization, with their mixing pattern imposed by the symmetries of the theory. The relation between the bare and renormalized operators is given by the renormalization mixing matrix $Z$, defined as:
\begin{equation}
  O_{(i)}^R= \sum_{j} \left(Z^{-1}\right)_{i j} O_{(j)}  
\end{equation}
where the indices $i$ and $j$ label the operators within a specific mixing set. The renormalization factors depend on both the regularization scheme $X$ (such as dimensional regularization (DR) or lattice regularization (LR)) and the renormalization scheme $Y$ (such as $\overline{\mathrm{MS}}$ or $\mathrm{RI}'$). These factors should be explicitly written as $Z^{X, Y}$, unless the context is unambiguous. To determine the mixing matrix elements $Z_{ij}$ on the lattice, we compute the $\rm \overline{MS}$ renormalized Green's functions in both dimensional and lattice regularization.

\subsection{Renormalization scheme}

Renormalization factors can be calculated nonperturbatively on the lattice within a suitably defined variant of the ${\rm RI}'$ scheme, unlike the $\overline{\rm MS}$ scheme whose definition is inherently perturbative. The transition to the $\overline{\rm MS}$ scheme is achieved using conversion factors between the two schemes. These conversion factors can only be determined through perturbation theory and are typically computed in DR rather than LR, as they are independent of the regularization scheme. They depend on the Wilson line length and the components of the renormalization-scale four-vector in the ${\rm RI}'$ scheme. Their definition is provided by:
\begin{equation}
    \mathcal{C}^{\rm \overline{MS}, {\rm RI}'} \equiv \left(Z^{\rm LR, \overline{MS}}\right)^{-1} \left(Z^{\rm LR, {\rm RI}'}\right) = 
    \left(Z^{\rm DR, \overline{MS}}\right)^{-1} \left(Z^{\rm DR, {\rm RI}'}\right) 
\end{equation}
Note that when mixing occurs among \(n\) operators, $\mathcal{C}^{\rm \overline{MS}, {\rm RI}'}$ are represented as an \(n \times n\) matrix.

\subsection{Lattice action}

We consider a nonabelian gauge theory of $SU(N_c)$ group and $N_f$ fermion multiplets, described by the following action:
\begin{equation}
    \hspace{-1cm}
    S=\frac{2}{g_0^2} \; \sum_{\rm plaq.} {\rm Re\,Tr\,}\{1-U_{\rm plaq.}\} + S_F
    \label{eq:action_nonlocal}
\end{equation}
The first term represents the gluonic contribution, where $U_{\rm plaq.}$ corresponds to the Wilson plaquette. The gluon field strength tensor $F_{\mu \nu}$ is obtained through the standard clover discretization while the Wilson line can be discretized on the lattice in a standard manner using gluon link variables. The second term of Eq.~\ref{eq:action_nonlocal} is the fermionic part of the action, \(S_F\), which contributes to the one-loop calculation only through the gluon field renormalization factor. We employ the clover-improved Wilson fermions \cite{Sheikholeslami1985}; however, adapting the results to other fermion actions at one-loop order is straightforward. 



\section{Results}

\subsection{Symmetries}
\label{sec:symmetries}

By studying the theory's symmetries, we can determine potential operator mixing under renormalization, as operators with the same transformations are typically susceptible to mixing.

Initially, we found that all nonlocal gluon operators that mix under renormalization will take the form of Eq.~(\ref{eq:nonlocal_operator}), with potentially different Lorentz indices $\mu,\nu,\rho,\sigma$. There is no mixing with other operators, such as nonlocal gluon operators with alternative Wilson line paths, nonlocal fermion operators, higher-dimensional operators, or non-gauge invariant operators (see \cite{Gavriel:2024qii} for further explanation).

The operators remain invariant under charge conjugation. Additionally, there are four "parity" transformations, corresponding to reflections $\mathcal{P}_1$, $\mathcal{P}_2$, $\mathcal{P}_3$, and $\mathcal{P}_4$ across each of the four axes in Euclidean space. Since some of these transformations change the sign of $z$, it is useful to change the basis of the nonlocal gluon operators, taking into account the translation invariance of the Lagrangian and the cyclic permutations of the trace in Eq.~(\ref{eq:nonlocal_operator}). Therefore, we define the following plus/minus basis, where these operators are now eigenstates of the parity transformations:
\begin{equation}
    O_{\mu \nu \rho \sigma}^{\pm} (z,0) 
    = \frac{1}{2} \left( O_{\mu \nu \rho \sigma} (z,0) \pm O_{\rho \sigma \mu \nu } (z,0) \right)
\end{equation}
We also study the rotational symmetry in the three dimensions perpendicular to the fixed direction of the Wilson line (rotational octahedral group. By combining our results from the parity transformation and the rotational analysis, we categorize the 36 operators into 16 groups, as shown in Table~\ref{tb:parity_rotation_symmetries}.

\begin{table}[ht]
\small
\begin{tabular}{c|c|c|c|c|c}
N. & Operators & N. & Operators & N. & Operators  \\[2pt] 
\hline
$\begin{array}{c}
1 \\ 2
\end{array}$ 
& 
$\begin{array}{c}
O_{3131}^+ +O_{3232}^+ +O_{3434}^+ \\
O_{1212}^+ +O_{1414}^+ +O_{2424}^+  
\end{array}$
&
9 
& 
$\left(\begin{array}{c}
O_{3212}^+ + O_{3414}^+ \\
O_{3121}^+ + O_{3424}^+ \\
O_{3141}^+ + O_{3242}^+
\end{array}\right)$
&
$\begin{array}{c}
13 \\
14
\end{array}$
& $\begin{array}{c}
            O_{3124}^+ +O_{3241}^+ +O_{3412}^+ \\
            O_{3124}^- +O_{3241}^- +O_{3412}^- 
            \end{array}$
\\
$\begin{array}{c}
3 \\ \\ 4
\end{array}$
& 
$\begin{array}{c}
\left(\begin{array}{c}
2 O_{3434}^+ -O_{3131}^+ -O_{3232}^+ \\
O_{3131}^+ -O_{3232}^+
\end{array}\right) \\ 
\left(\begin{array}{c}
2 O_{1212}^+ -O_{1414}^+ -O_{2424}^+ \\
O_{1414}^+ -O_{2424}^+
\end{array}\right)
\end{array}$ 
&
10 
& 
$\left(\begin{array}{c}
O_{3212}^- + O_{3414}^- \\
O_{3121}^- + O_{3424}^- \\
O_{3141}^- + O_{3242}^-
\end{array}\right)$  
&
$\begin{array}{c}
15 \\ \\
16
\end{array}$ & 
$\begin{array}{c}
\left(\begin{array}{c}
            2 O_{3412}^+ -O_{3241}^+ -O_{3124}^+ \\
            O_{3124}^+ -O_{3241}^+
            \end{array}\right) \\
\left(\begin{array}{c}
            2 O_{3412}^- -O_{3241}^- -O_{3124}^- \\
            O_{3124}^- -O_{3241}^-
            \end{array}\right)
\end{array}$
\\
\{5,6\} & $\left(\begin{array}{c}
            O_{3132}^- \\
            O_{3431}^- \\
            O_{3234}^-
            \end{array}\right)$  
            \; , \;  $\left(\begin{array}{c}
            O_{4142}^- \\
            O_{2421}^- \\
            O_{1214}^-
            \end{array}\right)$ 
&
11 & $\left(\begin{array}{c}
            O_{3212}^+ - O_{3414}^+ \\
            O_{3121}^+ - O_{3424}^+ \\
            O_{3141}^+ - O_{3242}^+
            \end{array}\right)$
& &
            \\
\{7,8\} & $\left(\begin{array}{c}
            O_{3132}^+ \\
            O_{3431}^+ \\
            O_{3234}^+
            \end{array}\right)$ 
            \; , \; 
            $\left(\begin{array}{c}
            O_{4142}^+ \\
            O_{2421}^+ \\
            O_{1214}^+
            \end{array}\right)$ 
&
12 & $\left(\begin{array}{c}
            O_{3212}^- - O_{3414}^- \\
            O_{3121}^- - O_{3424}^- \\
            O_{3141}^- - O_{3242}^-
            \end{array}\right)$    
& &
\end{tabular}
\caption{Groups of operators classified based on parity transformations and rotational symmetry. Operators in parentheses support 2- or 3-dimensional irreducible representations of the octahedral group.}
\label{tb:parity_rotation_symmetries}
\end{table} 

Operators in pairs $\{1, 2\}$, $\{3, 4\}$, $\{5, 6\}$, and $\{7, 8\}$ have the potential to mix under renormalization because they transform in the same way under parity and rotations. On the other hand, we find that operators in groups 9-16 cannot mix as they do not share the same transformation properties. It is worth noting that groups containing doublets or triplets have identical renormalization and mixing coefficients for each component within the multiplet.

We stress that all the conclusions drawn above, which rely solely on symmetry properties, hold true even beyond perturbation theory. Therefore, by employing appropriate operators listed in Table~\ref{tb:parity_rotation_symmetries}, one can eliminate mixing effects in numerical simulations.

\subsection{Perturbative Calculations}
\label{sec:results}

The one-particle-irreducible (1-PI) two-point bare amputated Green’s functions under consideration are:
\begin{equation}
    \delta^{(4)}(q+q') \ \Lambda_O (q,z) = \langle A^{a}_{\alpha} (q)\  \left(\int d^4 x \, O_{\mu \nu \rho \sigma}(x+z\hat\tau,x)\right) \ A^{b}_{\beta} (q')  \rangle_{\rm amp}
    \label{eq:Green_f_nonlocal}
\end{equation}
Here, $A^{a}_{\alpha}(q)\text{,} \ A^{b}_{\beta}(q')$ are two external gluon fields, and $O_{\mu \nu \rho \sigma}$ is the nonlocal gluon operator defined in Eq.~\ref{eq:nonlocal_operator}. Superscripts (subscripts) denote color (Lorentz) indices. To study the renormalization of the nonlocal gluon operators, we evaluate these Green's functions in both DR and LR. The calculations are carried out in a generic gauge, with off-shell gluons, and with general Lorentz indices for the external gluons and the operator.

We find that the renormalization function of the operators is diagonal when using the $\overline{\rm MS}$ renormalization condition and the one-loop results for the $\overline{\rm MS}$ renormalized Green's functions in DR. This holds both in the original basis ($O_{\mu \nu \rho \sigma}$) and in the basis provided in Table~\ref{tb:parity_rotation_symmetries}. In the latter, the diagonal matrix $Z^{\rm DR, \rm \overline{MS}}_{ij}$ is given by: 
\begin{equation}
    Z^{\text{DR}, \overline{\rm MS}}_{ij} = \delta_{ij} \left[ 1 + \frac{g^2}{16 \epsilon \pi^2} \left( \left( \frac{5}{3} + \omega_{i} \right) N_c - \frac{2}{3} N_f \right) \right], \quad \omega_{i} = \begin{cases} 
    0 & \text{for } i = 2,4,6,8 \\
    1 & \text{for } i = 9 \text{-}16 \\
    2 & \text{for } i = 1,3,5,7 
    \end{cases}
    \label{eq:renorm_factor_operator_new_basis}
\end{equation}
It is independent of the gauge parameter, as expected from gauge invariance in $\overline{\rm MS}$. Furthermore, Eq.~\ref{eq:renorm_factor_operator_new_basis} does not depend on the length of the Wilson line (\(z\)), since no dimensionless quantity involving \(z\) can appear in the pole part. This is expected to hold at all orders in perturbation theory. 

Note that the renormalization factor is identical operators which are members of the same multiple of the octahedral group.

\subsubsection{\texorpdfstring{${\rm RI}'$}{TEXT} renormalization prescription}
\label{subsec:RIprime}

The renormalization conditions for Green's functions in the ${\rm RI}'$ scheme can be defined in different ways, which may differ by finite terms. This is particularly relevant for operator mixing, where it is advantageous to consider the smallest possible set of operators that can mix under renormalization, as dictated by symmetries. In our case, this minimal set of nonlocal gluon operators consists of the groups $\{1,\ 2\}$, $\{3,\ 4\}$, $\{5,\ 6\}$, and $\{7,\ 8\}$, as shown in Table~\ref{tb:parity_rotation_symmetries}. Importantly, this choice must remain independent of the regularization method while accounting for any potential additional finite or power-divergent mixing, such as that encountered in lattice regularization.

Thus, we consider four $2\times2$ mixing matrices, corresponding to the four pairs of mixing operators, along with eight $1\times1$ matrices for operators that are multiplicatively renormalizable. An exception is operator 13, whose bare Green's function (Eq.~\ref{eq:Green_f_nonlocal}) vanishes, preventing its renormalization conditions from being directly defined. To address this, additional calculations would be necessary involving other Green's functions, such as three-point Green's functions. In total, 23 conditions are needed to determine the elements of the aforementioned matrices. This choice provides a suitable ${\rm RI}'$-like scheme useful for nonperturbative studies. 

\subsection{Conversion factors}

We studied the conversion factors between the $\overline{\rm MS}$ and $\rm RI'$ renormalization schemes for the groups of nonlocal gluon operators presented in Table~\ref{tb:parity_rotation_symmetries}. They can be easily derived by applying the renormalization conditions of the $\rm RI'$ scheme on the calculated $\overline{\rm MS}$-renormalized Green's functions. The $2 \times 2$ conversion factors for the operator mixing pairs are denoted as $\mathcal{C}_{{i,j}}^{\rm \overline{MS}, RI'}$, where $i$ and $j$ represent the operators in the mixing pair. For the multiplicatively renormalizable operators, a single index is used, as in $\mathcal{C}_{{i}}^{\rm \overline{MS}, RI'}$.

Due to the complexity and length of the conversion factors, rather than presenting the explicit expressions that involve integrals of modified Bessel functions of the second kind over a Feynman parameter, we present plots for the parameters commonly used in lattice simulations. Specifically, we have used the parameters of the $N_f = 2 + 1 + 1$ ensemble of twisted-mass clover-improved fermions described in Ref.~\cite{Delmar:2023agv}. In this lattice configuration, the Landau gauge is used ($\beta = 1$) with a lattice spacing of $a = 0.0938$ fm, and a lattice volume of $32^3 \times 64$ in lattice units. The $\overline{\rm MS}$ scale is fixed at $\overline{\mu} = 2 \ \text{GeV}$, while $g^2 = 3.47625$. In lattice units, the renormalization scale of ${\rm RI}'$ is defined as $a\bar{q} = \left( \frac{2\pi}{L} n_1, \frac{2\pi}{L} n_2, \frac{2\pi}{L} n_3, \frac{2\pi}{T} \left( n_4 + \frac{1}{2} \right) \right)$, where $n_i$ are integers. To ensure the antiperiodic boundary conditions are satisfied for the fermion fields in the temporal direction, we select isotropic spatial directions ($n_1 = n_2 = n_3$) when possible; further, we set a nonzero twist of $1/2$ in the temporal component. 

\begin{figure}[h]
    \centering
    \begin{subfigure}{0.32\linewidth}
        \centering
        \includegraphics[width=\linewidth]{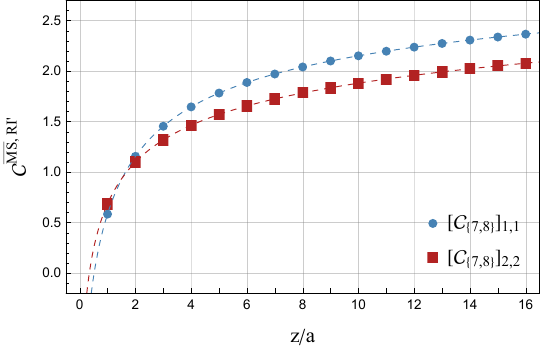}
        \caption{Diagonal elements of $\mathcal{C}_{\{7,8\}}^{\rm \overline{MS}, {\rm RI}'}$.}
    \end{subfigure}
    \hfill
    \begin{subfigure}{0.32\linewidth}
        \centering
        \includegraphics[width=\linewidth]{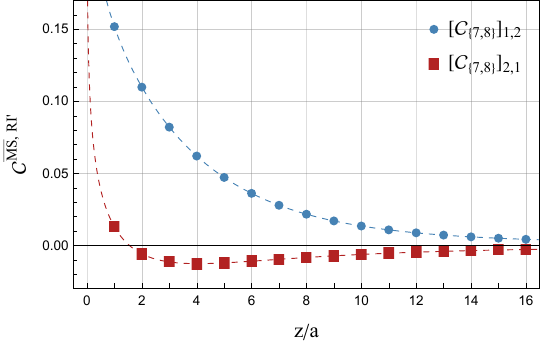}
        \caption{Nondiagonal elements of $\mathcal{C}_{\{7,8\}}^{\rm \overline{MS}, {\rm RI}'}$.}
    \end{subfigure}
    \hfill
    \begin{subfigure}{0.32\linewidth}
        \centering
        \includegraphics[width=\linewidth]{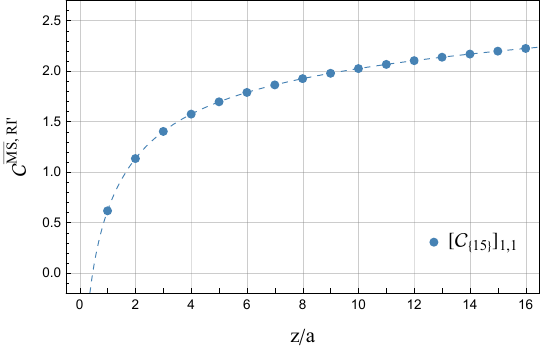}
        \caption{Conversion factor of $\mathcal{C}_{\{15\}}^{\rm \overline{MS}, {\rm RI}'}$. }
    \end{subfigure}
    \caption{Elements of conversion matrices of the plus-type operators as a function of $z/a$ with values $n_1 = n_2 = 3$, $n_3 = 0$, and $n_4 = -1/2$. Other plus-type mixing pairs and plus-type multiplicatively renormalized operators have similar qualitative behavior.}
    \label{fig:conversion_factors_plus_type}
\end{figure}

\begin{figure}[h]
    \centering
    \begin{subfigure}{0.32\linewidth}
        \centering
        \includegraphics[width=\linewidth]{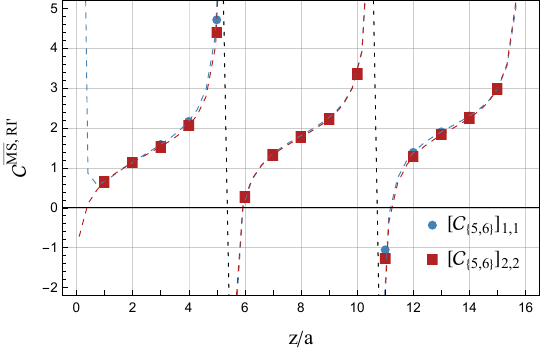}
        \caption{Diagonal elements of $\mathcal{C}_{\{5,6\}}^{\rm \overline{MS}, {\rm RI}'}$.}
    \end{subfigure}
    \hfill
    \begin{subfigure}{0.32\linewidth}
        \centering
        \includegraphics[width=\linewidth]{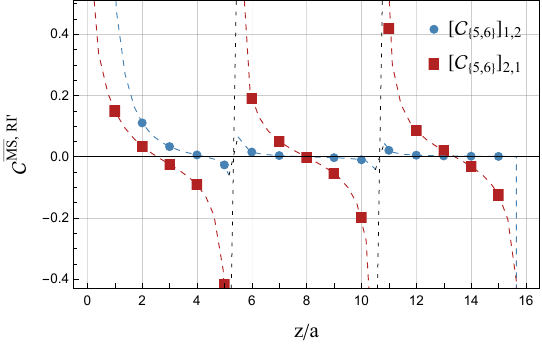}
        \caption{Nondiagonal elements of $\mathcal{C}_{\{5,6\}}^{\rm \overline{MS}, {\rm RI}'}$.}
    \end{subfigure}
    \hfill
    \begin{subfigure}{0.32\linewidth}
        \centering
        \includegraphics[width=\linewidth]{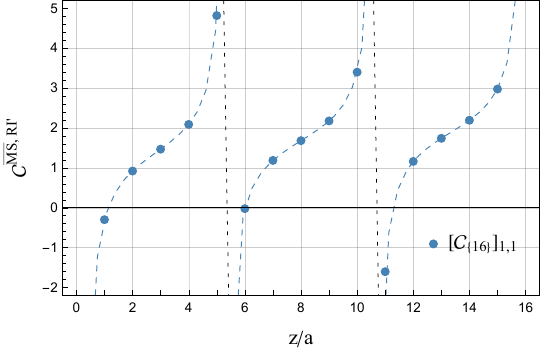}
        \caption{Conversion factor of $\mathcal{C}_{\{16\}}^{\rm \overline{MS}, {\rm RI}'}$.}
    \end{subfigure}
    \caption{Elements of conversion matrices of the minus-type operators as a function of $z/a$. The values $n_2 = n_3 = 3$, $n_1 = 0$, and $n_4 = 5$ are employed for the pair $\{5,6\}$ while for operator 16 we have selected $n_1 = n_2 = 0$, $n_3 = 3$, and $n_4 = 5$. The rest of the multiplicatively renormalizable minus-type operators follow the same format as operator 16.}
    \label{fig:conversion_factors_minus_type}
\end{figure}

As an example, to illustrate the behavior of the conversion matrix elements, we plot the elements for some of the plus-type operators (i.e., mixing pairs  $\{1,2\}$, $\{3,4\}$, and $\{7,8\}$ and multiplicatively renormalized operators 9,11, and 15) in Fig.\ref{fig:conversion_factors_plus_type}. Additionally, the conversion matrix elements for some minus-type operators (i.e., mixing pairs  $\{5,6\}$ and multiplicatively renormalized operators 10,12,14 and 16) are shown in Fig.\ref{fig:conversion_factors_minus_type}. 

The plots display only the real part of the conversion factors as a function of the Wilson line length rescaled by the lattice spacing (i.e., $z/a$). For the plus-type operators, the imaginary part is exactly zero, as a result of the chosen renormalization conditions, while the imaginary part for the minus-type operators is nearly zero, at most $10^{-5}$. 

The conversion factors for the plus-type and minus-type operators are plotted only in the positive direction of the Wilson line, specifically for positive values of $z$ up to half the lattice size. A singularity is expected at $z=0$, where the nonlocal operator becomes a local composite operator with additional contact singularities. As a result, we exclude the point $z/a=0$. Furthermore, we do not plot negative values of $z$ for the conversion matrix elements, as they are symmetric with respect to $z=0$ by definition and due to the chosen RI$'$ renormalization conditions.

The markers on the plots represent the values of the conversion factors at integer values of $z/a$ within the range from $1$ to $L/2=16$. The dashed lines connecting these markers indicate the values of the conversion factors for arbitrary, noninteger values of $z/a$. Note that the conversion factors can grow significantly depending on the chosen numerical values of $\bar{q}$. Therefore, it is essential to select these values appropriately to avoid this and to ensure that the tree-level Green's functions are invertible for all integer values of $z/a$ in the range $1\leq z/a < L/2$.

\subsection{Lattice Regularization}

Using lattice regularization, we now evaluate the bare Green's functions at one loop, as given by Eq.~(\ref{eq:Green_f_nonlocal}). The extraction of divergences from lattice integrals involves several subtleties, making this calculation significantly more complex than in DR. In addition to the nontrivial dependence on $z$ in the expressions, an overall factor of $1/a^2$ arises due to the presence of external gluons in the Green's functions. To eliminate this factor, we carefully extracted two powers of the external momentum $(aq)$. Furthermore, many diagrams have more complex tensorial structures compared to the tree level. These involve integrals over Bessel functions that depend on the momentum of the Green's function.

From these calculations, we find that the renormalization factors of the nonlocal gluon operators are in diagonal form at one-loop calculations. This is true in both the original basis ($O_{\mu \nu \rho \sigma}$) and the basis shown in Table~\ref{tb:parity_rotation_symmetries}, as in DR. Therefore, the nonlocal gluon operators under study at one loop level in lattice theory are multiplicatively renormalized. In the basis of Table~\ref{tb:parity_rotation_symmetries},  the matrix $Z^{\rm LR, \rm \overline{MS}}_{ij}$ can be expressed as:
\begin{align}
    Z^{\text{LR}, \overline{\rm MS}}_{ij}= \delta_{ij} \Bigg[ 1+ \frac{g^2}{16 \pi^2} \Bigg\{ &\frac{2 \pi^2}{N_c} + N_f \left(e_1 + e_2\, c_{SW} + e_3\, c_{SW}^2 + \frac{2}{3} \log(a^2 \bar{\mu}^2 ) \right) \nonumber \\ 
    &+ N_c \left(e_4 + e_5 \frac{|z|}{a} - \frac{5}{3} \log(a^2 \bar{\mu}^2 ) - \left( e_6 + \log(a^2 \bar{\mu}^2 ) \right) \omega_{i} \right)  \Bigg\}  \Bigg]
    \label{eq:renorm_factor_operator_lattice_new_basis}
\end{align}
Here, $\omega_{i}$  is defined by Eq.~\ref{eq:renorm_factor_operator_new_basis}, and the coefficient values are $e_1=-1.05739$, $e_2=0.79694$, $e_3=-4.71269$, $e_4=-17.81504$, $e_5=-19.95484$, and $e_6=-8.37940$. Furthermore, the above equation is independent of the gauge parameter. The accuracy of the numerical loop integration is found to be up to $\mathcal{O}(10^{-5})$, as indicated by the precision of the coefficients and the cancellation of gauge dependence. Note that the $c_{SW}$ term arises because, in the renormalized Green's function, one must account for the gluon field renormalization function.

As in the case of nonlocal fermion operators, there is a linear divergence that depends on the length of the Wilson line, with its coefficient $|e_5|$ matching the corresponding divergent term in the quark nonlocal operators for an arbitrary Wilson line shape~\cite{Spanoudes:2024kpb}. This occurs because the linear divergence arises solely from Wilson-line self-energy diagrams. There are also logarithmic divergences as expected which arise from the endpoints and contact points of the Wilson lines.

\section{Conclusion}

First, based on the symmetries of the theory, we categorize the nonlocal gluon operators into the groups shown in Table~\ref{tb:parity_rotation_symmetries}. We find that the operator pairs $\{1,\ 2\}$, $\{3,\ 4\}$, $\{5,\ 6\}$, $\{7,\ 8\}$ mix under renormalization, while the remaining operators, 9–16, are multiplicatively renormalizable.

We evaluated the renormalization factors in the $\rm \overline{MS}$ scheme using both dimensional and lattice regularization at the one-loop level. In the continuum, these factors are diagonal, consistent with previous studies employing the auxiliary-field formulation~\cite{Dorn:1981wa,Zhang2018,Braun:2020ymy}. On the lattice, the renormalization factors also exhibit a diagonal form. According to Eq.~\ref{eq:renorm_factor_operator_lattice_new_basis}, all the nonlocal gluon operators under consideration undergo multiplicative renormalization. However, as indicated by the symmetry properties of the theory, we anticipate that mixing among operator pairs will emerge at higher orders. Furthermore, the conversion factors of the nonlocal gluon operators between the $\rm \overline{MS}$ renormalization scheme and the RI$'$ scheme are calculated, with the RI$'$ scheme being defined as suitable for nonperturbative studies and compatible with the mixing pattern of the operators.
 

By providing insights into the renormalization of nonlocal gluon operators, we expect that the results of this study will contribute significantly to the investigation of gluon PDFs through lattice QCD. 

\begin{acknowledgments}
D.G., H.P., and G.S. acknowledge financial support from the projects EXCELLENCE/0421/0025 and CONCEPT/0823/0052, implemented under the "THALIA 2021-2027" program, co-funded by the EU and the Research and Innovation Foundation (RIF). G.S. also acknowledges funding from the 3D-nucleon project (EXCELLENCE/0421/0043), co-financed by the EU and the Republic of Cyprus through RIF.
\end{acknowledgments}

\bibliographystyle{JHEP}
\bibliography{references}

\end{document}